# Implementation of large momentum transfer without swapping the directions of the Raman beams

Jinyang Li[1], Jason Bonacum[2] and Selim M Shahriar[1,3]

[1]*Department of Physics and Astronomy, Northwestern University, Evanston, IL 60208, USA*

[2]*Digital Optics Technologies, Rolling Meadows, IL 60008 USA*

[3]*Department of Electrical and Computer Engineering, Northwestern University, Evanston, IL 60208, USA*

**Abstract**

Large momentum transfer (LMT) is an important technique for magnifying the phase shift accumulated in an atom interferometer. Existing approaches to implement Raman-transition-based LMT all involve physically swapping the propagation directions of the two counterpropagating Raman beams repeatedly, which could significantly complicate the experimental system. Here, we demonstrate a simpler approach for Raman-transition-based LMT that does not involve a physical swap of the directions of the Raman beams. In this approach, both Raman beams are retroreflected, and a Doppler shift induced by a bias velocity of the atoms is used to separate the transition frequencies of the two pairs of counterpropagating Raman beams. Therefore, an effective swap of the directions of the Raman beams can be achieved by shifting the relative frequency between the two Raman beams from the resonant frequency of one pair of the Raman beams to that of the other pair. We demonstrate the use of this technique for LMT-augmented accelerometry using atoms released from a magneto-optic trap.

## 1. Introduction

Light-pulse atom interferometry is a preeminent method for inertial measurement [1, 2], which can be used for inertial navigation [3, 4, 5] and fundamental science including gravitational wave detection [6] and equivalence principle test [7, 8, 9]. Large momentum transfer (LMT) [10, 11, 12, 13, 14, 15, 16, 17] is a technique widely used to magnify the phase shift in an atom interferometer, which can potentially increase the measurement sensitivity. LMT has been applied to single-photon-transition-based interferometers [10, 11], Bragg-transition-based interferometers [12, 13], and Raman-transition-based interferometers [14, 15, 16, 17]. For alkaline atoms, only the last two types of transitions apply. The Bragg transition has the advantage of low relative light shift, while has the disadvantage of requiring very cold atoms with a Doppler broadening less than tens of kHz. To achieve such a temperature, evaporative cooling or velocity selection is needed, either of which significantly reduce the number of atoms for interrogation [17]. Therefore, it is of interest to investigate LMT for Raman-transition-based interferometers employing atoms that have been released from a magneto-optic trap and cooled further only via optical molasses.

In a conventional Raman-transition-based interferometer, a temporal sequence of $\pi/2$, $\pi$, $\pi/2$ Raman pulses are applied. For convenience, the effective direction of a pair of counterpropagating Raman beams is defined as the propagation direction of the higher-frequency Raman beam. For the first order LMT, four additional $\pi$ Raman pulses are applied with the effective direction antiparallel to that of the original three pulses. For *n*-order LMT, 4*n* additional $\pi$ Raman pulses are needed. We have shown previously [14] that the LMT process can be applied to enhance the sensitivity of a point-source interferometer as well. In that case, we found that the optimal value of the LMT order depends on the choice of experimental parameters. Specifically, we showed that the value of the optimal order can be as high as 60 for experimental accessible experimental conditions. For such a system, one would need 240 additional $\pi$ Raman pulses with alternating direction. For such a scenario, a key challenge of implementing LMT is swapping the propagation directions of the two Raman beams during the Raman pulse sequence. One possible method used to exchange the directions of the two Raman beams is to swap the polarizations of the two beams before a polarizing beam splitter (PBS) [16, 17]. This approach has several drawbacks. First, directing the two Raman beams along two paths increases the complexity of the system significantly. Second, in this approach, the two Raman beams must have orthogonal polarizations and thus cannot be two components in the output from an electro-optic modulator (EOM). To generate Raman beams with orthogonal polarizations for alkaline atoms like $^{87}$Rb and $^{133}$Cs, use of acousto-optic modulators (AOMs) is not feasible due to the large energy gap between the two hyperfine ground states. There are two alternative methods. One is to off-set phase-lock a diode laser to another laser that produces one Raman beam, so that the output from this diode laser can work as the other Raman beam. However, for diode lasers used for CS or Rb, the phase lock does not produce a beat note with very low phase noise level, which is manifested by the wide pedestal around the central peak in the beat note spectrum [18]. The other alternative is to split a beam into two parts, one passing through an EOM and the other unaffected. When the carrier in the output of the EOM is suppressed, the first order frequency component and the unaffected beam can work as the Raman beams. The requirement to suppress the fundamental completely makes this approach highly challenging. Third, to change the polarizations of the two beams rapidly without wasting optical power, a Pockels cell needs to be used [16]. A Pockels cell has shortcomings, such as limited output optical power, and undesirable distortion of the beam polarization. Furthermore, it requires the use of a high-voltage power supply, which can be problematic in situations where minimizing the SWaP (size, weight and power) of the device may be important.

To overcome the complexities brought about by the physical swap of the propagation directions of the Raman beams, we have demonstrated a simpler approach for LMT. In this approach, both Raman beams are retroreflected. In this way, there are two pairs of counterpropagating Raman beams in opposite effective directions. A bias velocity of the atoms generates a Doppler shift that separates the two-photon resonance frequencies for the two pairs of Raman beams. Therefore, an effective swap of the directions of the Raman beams can be achieved by shifting the relative frequency between the two Raman beams from the resonant frequency of one pair of the Raman beams to that of the other pair [19]. It should be noted that the process would work for bias velocities that could possibly change with time, as long as the minimum bias velocity is large enough to ensure there is no overlap between the spectral profiles of the Raman transitions induced by the two pairs of Raman beams. The widths of the spectral profiles depend on factors such as the Raman Rabi frequency for each pair as well as the temperature of the atoms. To impart a velocity in an arbitrary direction, the launching technique [19] can be used. For the experiment described here, the bias velocity is due to gravitational acceleration, with the initial fall time (prior to the application of the first Raman pulse) chosen to be large enough to produce a minimum bias velocity that satisfies the constraint mentioned above.

The rest of the paper is organized as follows. In Sec. 2, we describe the experimental system for the atom interferometer. In Sec. 3, we explain the approach used for implementing LMT. In Sec. 4, we present experimental results. The conclusion and discussion are presented in Sec. 5.

## 2. Experimental system

In our atom interferometer, $^{85}$Rb atoms are manipulated in a glass vacuum cell with a cross section of $3 \text{ cm} \times 3 \text{ cm}$. The optical system is shown in Figure 1. An external cavity diode laser (ECDL, model: TAPro) is locked to the $S_{1/2}, F = 3 \rightarrow P_{3/2}, F = 2, 4$ cross-over resonance using saturated absorption spectroscopy. In this way, the ECDL is 92 MHz red-detuned from the $S_{1/2}, F = 3 \rightarrow P_{3/2}, F = 4$ transition. The beam for the magneto-optic trap (MOT) is produced by an AOM with a center frequency of 40 MHz set in a cats'-eye double-pass configuration. For operating the MOT, the AOM is modulated at 40 MHz, which produces a net upshift of the ferquency by 80 MHz. Therefore, the MOT beam is 12 MHz red-detuned from the $S_{1/2}, F = 3 \rightarrow P_{3/2}, F = 4$ transition. During the polarization gradient cooling (PGC) stage, the modulation frequency for the AOM is ramped down to 20 MHz, which increases the detuning of the MOT beam to 52 MHz, while reducing the MOT beam intensity by a factor of seven. The cats'-eye double-pass configuration prevents mialignment of the beam during this change in the modulation frequency. The probe beam is resonant with the $S_{1/2}, F = 3 \rightarrow P_{3/2}, F = 4$ transition and is produced by shifting a piece of the output from the ECDL by 92 MHz with another AOM. A distributed Bragg reflector laser (DBR 1, model: Photodigm 780.241DBRH) is locked to the ECDL with an offset phase lock loop (OPLL, model: Vescent D2-135). The frequency of DBR 1 is 1 GHz lower than that of the ECDL. The output from DBR 1 is amplified by a tapered amplifier (TA, model: Sacher TEC-400-0780-2500) and is then coupled into a fiber-couple EOM (model: iXblue NIR-MPX800-LN-20) driven at a frequency of about 3.0 GHz. The carrier and the positive first order frequency component stimulate the Raman transition. An AOM is set before the EOM to work as a fast switch. Another DBR laser (DBR 2, model: Photodigm 780.241DBRH) is locked to the $S_{1/2}, F = 2 \rightarrow P_{3/2}, F = 2,3$ cross-over resonance. However, since the peak of the $S_{1/2}, F = 2 \rightarrow P_{3/2}, F = 2,3$ cross-over resonance and the peak of the $S_{1/2}, F = 2 \rightarrow P_{3/2}, F = 1,3$ cross-over resonance cannot be well resolved in the saturated absorption spectrum, DBR 2 is roughly 40 MHz red-detuned from the $S_{1/2}, F = 2 \rightarrow P_{3/2}, F = 3$ transition. Therefore, the repump beam is produced by shifting up the output of DBR 2 by 40 MHz with another AOM.

All these beams are coupled into polarization-maintaining fibers and directed to the vacuum cell. The intensity of the single-axis MOT beam applied to the vacuum cell is about 14 $mW \cdot cm^{-2}$ at the MOT and is reduced to about 2 $mW \cdot cm^{-2}$ during the PGC stage. The output from the EOM is turned into being $\sigma^+$-polarized with a quarterwaveplate and there is no further polarization manipulation on the beam. Therefore, all Raman beams are $\sigma^+$ polarized. The intensity of the output from the EOM (including all frequency components) is about 60 $mW \cdot cm^{-2}$. It should be noted that this intensity can cause photorefractive damage to the EOM if the input is continous. However, we use Raman pulses as short as tens of μs with a duty cycle of ~10% and long time sepration between repetition of the interferometer sequence. As scuh, the average power seen by the EOM remains below the photorefractive damage threshold [20].

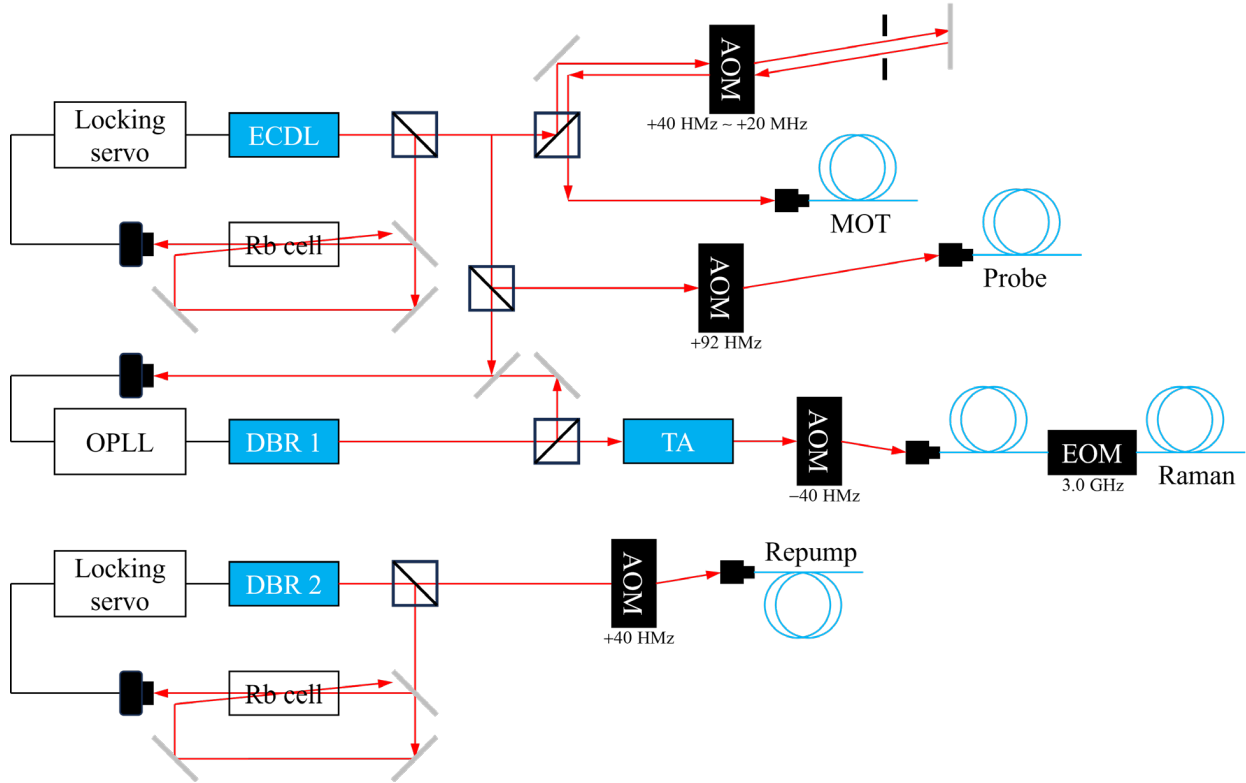

Figure 1. Optical system for generating the beams necessary for the atom interferometer. One ECDL, two DBR lasers, and a TA are used in this system. AOMs are used in this system to shift the beam frequencies to the required values and to work as fast switches to the beams. The fiber-coupled EOM is used to produce the Raman beams. All beams are coupled into polarization maintaining fibers and directed to the vacuum cell.

The frequencies of the beams are shown in Figure 2(a) and the pulse sequence is shown in Figure 2(b). The steps of the experiment are as follows. First, about $10^7$ atoms are captured in the MOT within 300 ms. What follows is the polarization gradient cooling step. To initiate this step, the magnetic quadrupole for the MOT is turned off and the modulation frequency for the AOM controlling the MOT beam is ramped down to 20 MHz within 5 ms, as discussed earlier. The atoms reach a temperature of about 14 μK after this step. As the cooling process ends, the MOT beam remains on for another 3 ms after the repump beam is turned off , in order to pump all atoms to $S_{1/2}, F = 2$ hyperfine state. A vertical bias magnetic field with a magnitude of about 3 G is applied along the direction of propagation of the Raman beams before the Raman pulse sequence to break the degeneracy of the Zeeman substates. Then the Raman pulse sequence is applied to induce the interference. Although the atoms are distributed among all Zeeman substates of the $F = 2$ hyperfine state, the relative frequency between the Raman beams together with their polarization configuration ensures that only the Raman transition between the two $m_F = 0$ Zeeman substates is stimulated. Finally, the probe beam is applied and the fluorescence emitted by the $F=3$ atoms is detected by a photodetector.

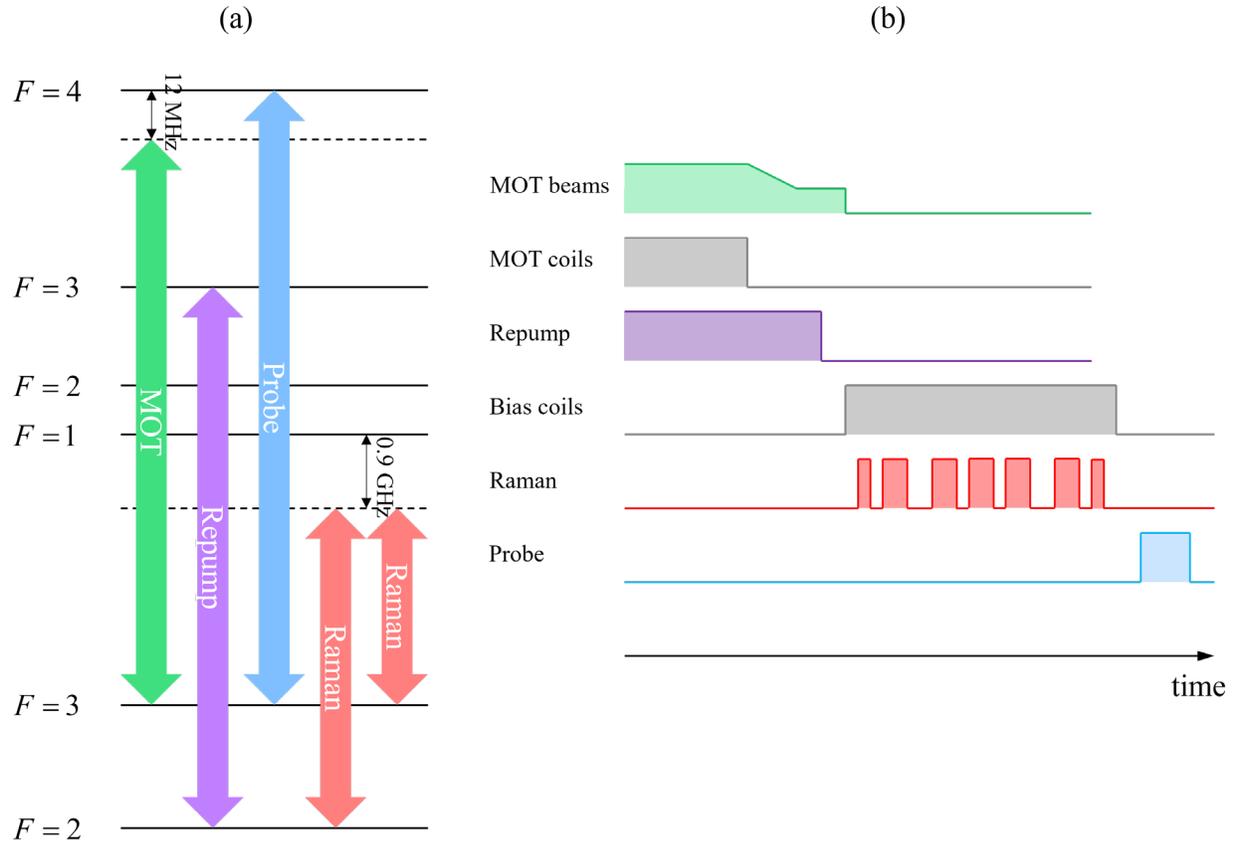

Figure 2. (a) Frequencies of the optical beams for the atom interferometer. (b) Pulse sequence of the atom interferometer. The pulses include optical pulses and electronic pulses for controlling the magnetic fields.

## 3. Implementation of LMT

We next describe our approach for LMT. After the light field for PGC is turned off, the atoms undergo a free fall for 20 ms and gain a velocity of $19.6 \text{ cm}\cdot\text{s}^{-1}$. Therefore, the transition frequencies of a pair of counterpropagating Raman beams are shifted by about $k_{\text{eff}} v = 2\pi \times 506 \text{ kHz}$, where $k_{\text{eff}}$ is the effective wavenumber of the counterpropagating Raman beams, which is about $2\pi \times 2.56 \times 10^4 \text{ cm}^{-1}$ for $^{85}\text{Rb}$, and $v$ is the bias velocity of the atoms along the Raman beams. However, the Doppler shifts for the two pairs of counterpropagating Raman beams have opposite signs. The shifted transition frequencies of the two pairs of Raman beams can be seen from the Raman spectrum, which is obtained by applying only one $\pi$ pulse during the Raman pulse sequence. An example of such a Raman spectrum is shown in Figure 3. There are two wider side peaks and a narrower central peak in the Raman spectrum. The two wider peaks correspond to the two pairs of counterpropagating Raman beams. The larger widths of the peaks result from the Doppler broadening. The narrower peak is produced by the copropagating Raman beams. Consequently, by shifting the relative frequency between the two Raman beams, namely the modulation frequency for the EOM, from the left peak to the right peak, the directions of the two Raman beams are effectively swapped, as shown in Figure 3.

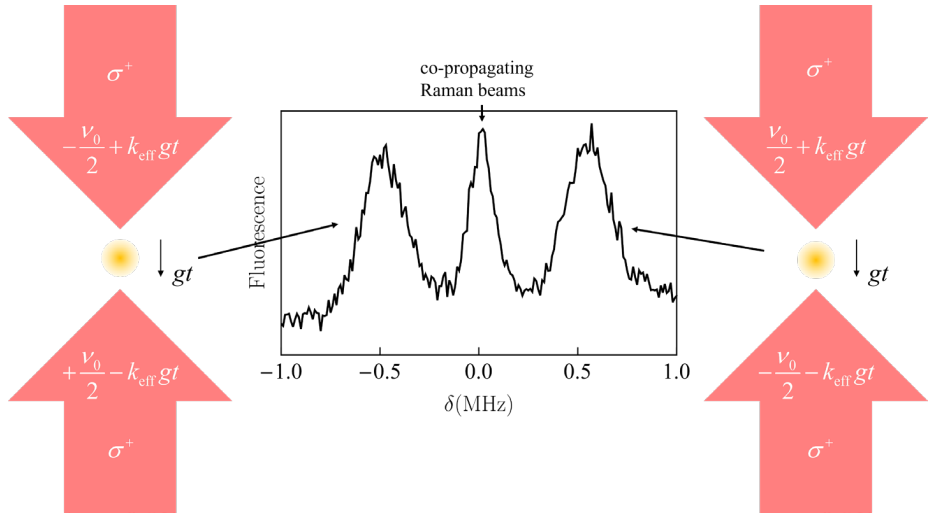

Figure 3. Raman spectrum and the Raman beams producing the peaks. The central narrow peak is produced by copropagating Raman beams, and the two broader peaks on the two sides are produced by two pairs of counterpropagating Raman beams in opposite effective directions. Here, $v_o$ is the hyperfine splitting frequency in the ground state.

The Raman pulse sequence for the first order LMT is shown in Figure 4(a). The corresponding modulation frequency for the EOM is shown in Figure 4(b). It should be noted that for the Raman spectrum, the transitions are from $|F=2, 0\hbar k_{eff}\rangle$ to $|F=3, \pm\hbar k_{eff}\rangle$, where the first part in the ket indicates the internal state and the second part the center-of-mass (COM) momentum state. The Zeeman substates are not shown in the kets since the atoms are always in the $m_F = 0$ Zeeman substates. In an interferometer augmented with LMT, however, the transitions can happen between many COM momentum states that differ by $\hbar k_{eff}$. The frequencies chosen for each pair of Raman beams have to be optimized to take this into account. The optimal values for the LMT are shown in Figure 4(b).

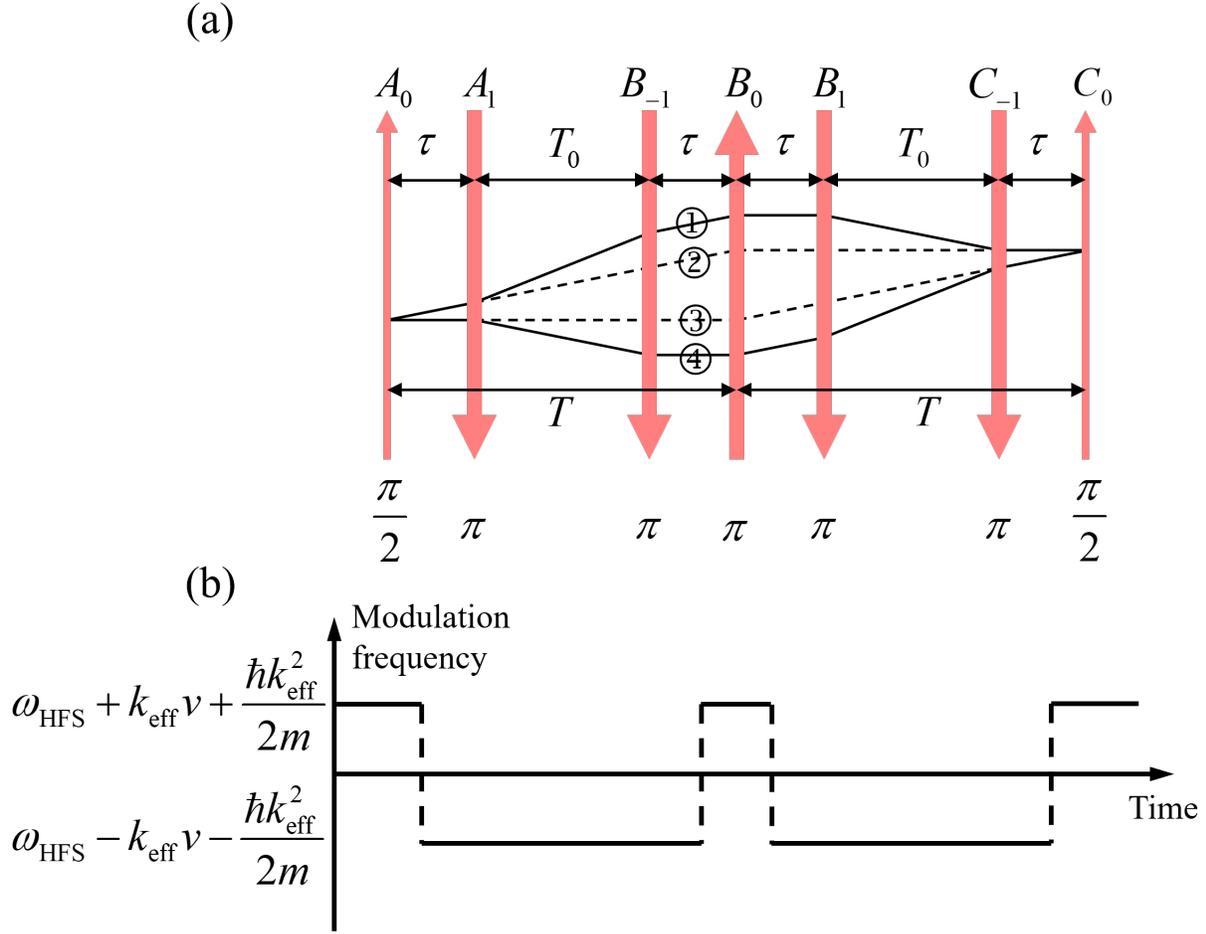

Figure 4. (a) Raman pulse sequence for the first order LMT. (b) Modulation frequency for the EOM as a function of time. For the three original pulses of an atom interferometer $A_0$, $B_0$, and $C_0$, the modulation frequency is $\omega_{\text{HFS}} + k_{\text{eff}} v + \hbar k_{\text{eff}}^2/2m$, while for the four additional pulses for LMT, the modulation frequency is $\omega_{\text{HFS}} - k_{\text{eff}} v - \hbar k_{\text{eff}}^2/2m$.

## 4. Experimental results

We first tested our atom interferometer without LMT. The signal as a function of $T$ is shown in Figure 5, where $T$ is the half duration of the Raman pulse sequence, as defined in Figure 4(a). The upper plot shows the case where the modulation frequency remains constant and lower plot shows the case where the modulation frequency is chirped to simulate an acceleration [21]. Here, the blue dots represent experimental data, and the red lines show theoretical fits, where the amplitudes and baselines of the fringes are used as free parameters. The important aspect of this data is to note the excellent agreement of the signal as a function of $T$.

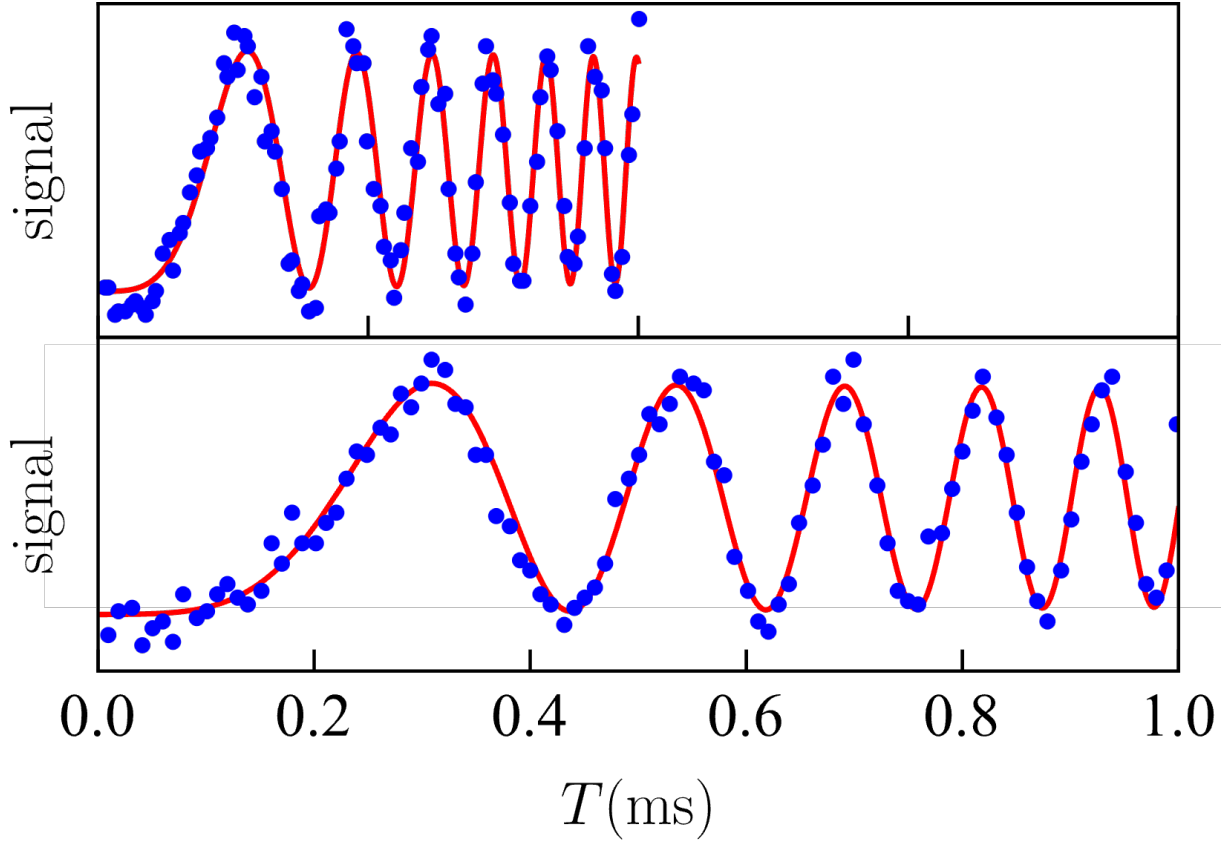

Figure 5. Signal of the bare atom interferometer as a function of $T$. The upper row shows the case where the modulation frequency for the EOM is held constant, and the lower row shows the case where the modulation frequency is chirped to simulate an acceleration that reduces the total effective acceleration by a factor of five.

We next tested the interferometer augmented with the first order LMT. It turns out that in addition to the desirable trajectories ① and ④ in Figure 4(a), a fraction of atoms also follow trajectories ② and ③ due to the approximately 50% transition efficiency of the $\pi$ pulses. Therefore, to see whether LMT really happens, we scanned $T_0$ while holding $T$ constant at 400 μs. The signal as a function of $T_0$ is shown in Figure 6. Due to the interference between trajectories ① and ③ as well as between trajectories ② and ④, in addition to the desirable interference between trajectories ① and ④, the signal should be in the form of $c_0 + c_1 \cos(k_{\text{eff}} g T T_0) + c_2 \cos(2 k_{\text{eff}} g T T_0)$. The solid fitting curve shown in Figure 6 agrees with this expression, with $c_2/c_1 = 0.44$. While the signal is rather noisy, the two dips are quite clear, indicating that the process of first order LMT is indeed occurring. The low signal-to-noise ratio is due to a set of imperfections in our current apparatus. Efforts are in progress to suppress these imperfections in order to be able to demonstrate higher order LMT processes.

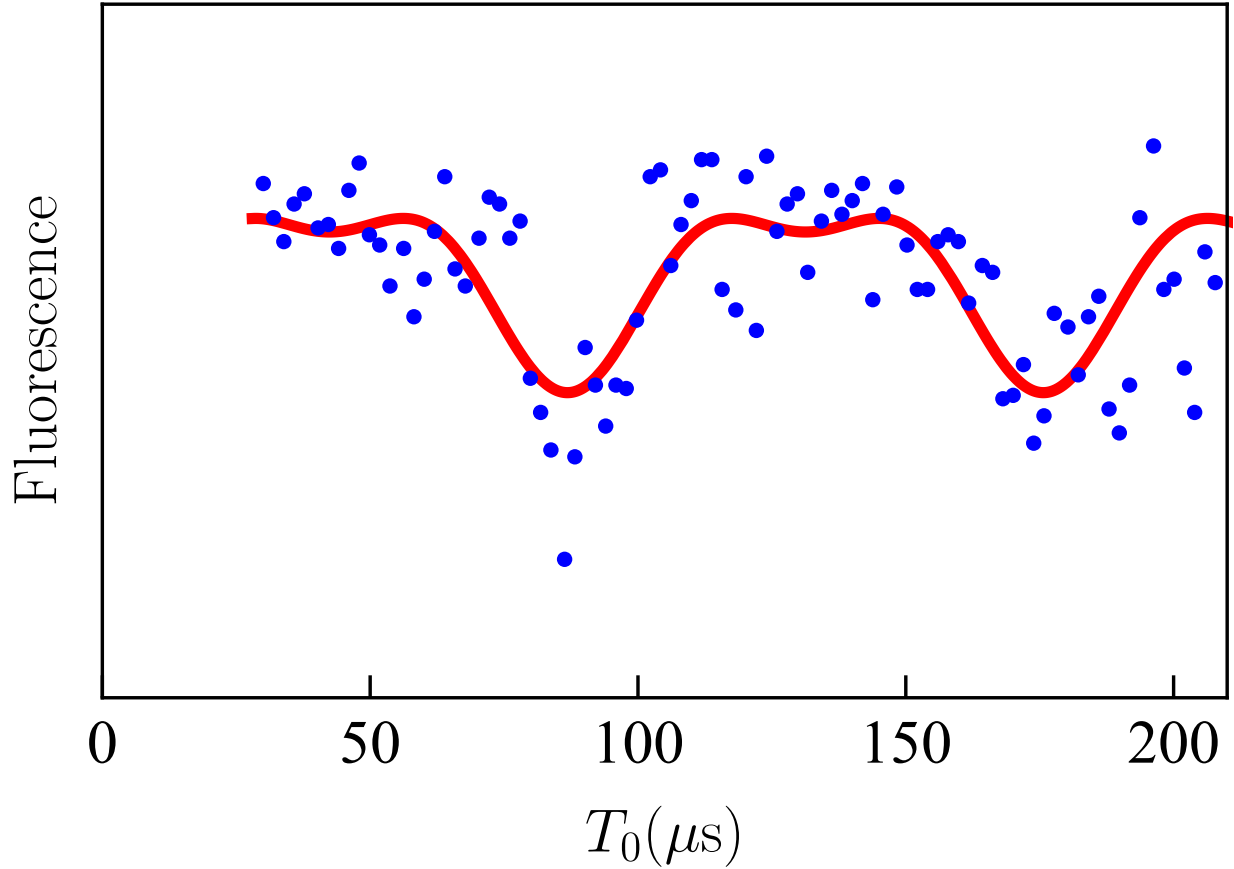

Figure 6 Signal of the atom interferometer augmented with the first order LMT as a function of $T_0$ when $T$ is held constant at 400 μs. The signal is in the form of $c_0 + c_1 \cos(k_{\text{eff}} g T T_0) + c_2 \cos(2k_{\text{eff}} g T T_0)$, with $c_2/c_1 = 0.44$.

## 5. Conclusion and discussion

We have demonstrated a simpler approach for Raman-transition-based LMT that does not involve a physical swap of the directions of the Raman beams. In this approach, both Raman beams are retroreflected, and a Doppler shift induced by a bias velocity of the atoms is used to separate the transition frequencies of the two pairs of counterpropagating Raman beams. Therefore, an effective swap of the directions of the Raman beams can be achieved by shifting the relative frequency between the two Raman beams from the resonant frequency of one pair of the Raman beams to that of the other pair.

The quality of our signal is limited by the transition efficiency of the Raman pulses. The transition efficiency could be enhanced by improving the degree of cooling, as well as increasing the diameter and the intensity of the Raman beams. In our experiment, the order of LMT is only one, which does not cause a large relative Doppler shift between the two arms of the interferometer.

For higher order LMT, it is important to address the relative Doppler shift between the two arms, especially when the relative Doppler shift is comparable to or higher than the Doppler broadening. A straightforward approach for addressing this problem is to modulate the EOM with a superposition of two frequency components, with each frequency component addressing one arm. However, this technique is applicable only if the relative Doppler shift significantly exceeds the Doppler broadening. An alternative technique that does not have this problem is the application of Floquet pulses [11]. This technique has only been applied to single-photon transitions so far. In principle, it can also be applied to Raman transitions.

**Funding.** National Aeronautics and Space Administration (80NSSC20C0161) and U.S. Department of Defense (W911NF202076).

**Disclosures.** The authors declare no conflicts of interests.

**Data availability.** Data underlying the results presented in this paper are not publicly available at this time but may be obtained from the authors upon reasonable request.